\documentclass[11pt]{article}

\usepackage{amsmath,amssymb,amsfonts}
\usepackage{amsthm}
\usepackage{natbib}
\usepackage{geometry}
\usepackage{setspace}
\usepackage{algorithm, algorithmic}  
\usepackage{mathtools}  
\usepackage[textwidth=8em,textsize=small]{todonotes}
\usepackage{enumerate}
\usepackage{url} 
\usepackage{xcolor}

\newtheorem{theorem}{Theorem}

\newtheorem{proposition}{Proposition}

\theoremstyle{definition}

\newtheorem{remark}{Remark}

\def\bSig\mathbf{\Sigma}

\def\pr{\hbox{pr}}
\def\hat{\widehat}

\def\T{{ \mathrm{\scriptscriptstyle T} }}

\def\A{{\bf A}}

\def\B{{\bf B}}

\def\f{{\bf f}}

\def\g{{\bf g}}

\def\V{{\bf V}}

\def\s{{\bf s}}

\def\X{{\bf X}}
\def\x{{\bf x}}

\def\W{{\bf W}}

\def\bt{{\boldsymbol\theta}}

\def\bb{{\boldsymbol\beta}}

\def\bOmega{{\boldsymbol\Omega}}
\def\bSigma{{\boldsymbol\Sigma}}
\def\bGamma{{\boldsymbol\Gamma}}

\def\0{{\bf 0}}

\def\pr{\hbox{pr}}

\def\log{{\rm log}}

\def\squarebox#1{\hbox to #1{\hfill\vbox to #1{\vfill}}}

\def\bse{\begin{eqnarray*}}
\def\ese{\end{eqnarray*}}
\def\be{\begin{eqnarray}}
\def\ee{\end{eqnarray}}
\def\bsq{\begin{equation*}}
\def\esq{\end{equation*}}
\def\bq{\begin{equation}}
\def\eq{\end{equation}}

\def\sumIP1{\sum_{i=1, i\in P_1}^N}

\title{\bf A Communication-Efficient
Distributed Algorithm for Learning with Heterogeneous and Structurally Incomplete Multi-Site Data}
\author{Xiaokang Liu\thanks{Co-first author. Department of Statistics and Data Science, University of Missouri} 
\and  Yuchen Yang \thanks{Co-first author. Department of Biostatistics, Epidemiology and Informatics, University of Pennsylvania} 
\and Yifei Sun\thanks{Department of Biostatistics,
       Columbia University} 
\and Jiang Bian\thanks{Department of Biostatistics and Health Data Science,
       Indiana University}
\and Yanyuan Ma\thanks{Department of Statistics, 
       The Pennsylvania State University} 
\and Raymond J. Carroll\thanks{Department of Statistics, 
       Texas A\&M University}
\and Yong Chen\thanks{Corresponding author. Department of Biostatistics, Epidemiology and Informatics, University of Pennsylvania, ychen123@upenn.edu}}
\date{}

\begin{document}
\maketitle

\begin{abstract}
In multicenter biomedical research, integrating data from multiple decentralized sites provides more robust and generalizable findings due to its larger sample size and the ability to account for the between-site heterogeneity.  However, sharing individual-level data across sites is often difficult due to patient privacy concerns and regulatory restrictions. To overcome this challenge, many distributed algorithms, that fit a global model by only communicating aggregated information across sites, have been proposed. A major challenge in applying existing distributed algorithms to real-world data is that their validity often relies on the assumption that data across sites are independently and identically distributed, which is frequently violated in practice. In biomedical applications, data distributions across clinical sites can be heterogeneous. Additionally, the set of covariates available at each site may vary due to different data collection protocols. We propose a distributed inference framework for data integration in the presence of both distribution heterogeneity and data structural heterogeneity. By modeling heterogeneous and structurally missing data using density-tilted generalized method of moments, we developed a general aggregated data-based distributed algorithm that is communication-efficient and heterogeneity-aware. We establish the asymptotic properties of our estimator and demonstrate the validity of our method via simulation studies. 
\end{abstract}


\newpage

\section{Introduction}
\label{sec:Intro}



Biomedical research with data from multiple decentralized sites are becoming popular recently. Integrating data from multiple clinical sites can provide more accurate evidence to facilitate clinical research and lead to more generalizable findings. However, due to the regulatory restrictions, sharing individual-level patient data across sites is often prohibited and often only aggregated data can be shared. To overcome this challenge, distributed algorithms, also known as ``divide-and-conquer" procedures, federated learning, or collaborative learning, are in great need. Distributed algorithms fit a model across multiple decentralized sites without exchanging individual-level patient data but only aggregated statistics or estimated model parameters \citep{zhang2012communication, battey2018distributed, jordan2018communication}. Among distributed algorithms, communication-efficient algorithms are particularly appealing as they only require few rounds of communication of aggregated data across sites. For example, \cite{wang2017efficient} and \cite{jordan2018communication} proposed a surrogate likelihood approach that approximates the global likelihood using the likelihood function at a local site and aggregated data from collaborating sites. This method has low communication cost and improves the performance over traditional meta-analysis \citep{duan2020learning}.

Despite the promise of the existing framework of distributed inference, there are still two major challenges when applying them to the real-world settings. The first is between-site population distribution heterogeneity. In \cite{wang2017efficient} and \cite{jordan2018communication}, the data at different sites are assumed to be independently and identically distributed. However, in real-world settings, the data across clinical sites are rarely homogeneous due to intrinsic differences in patient characteristics, clinical practices, and data quality control processes. For example, the OneFlorida network data contains longitudinal patient-level linked electronic health records (EHRs) and claims data for 15 million Floridians from 12 unique healthcare organizations across the state with diverse patient populations in terms of geographic setting (urban vs. rural), age, race, and ethnicity \citep{shenkman2018oneflorida}. University of Florida Health serves mostly non-Hispanic Whites and a mixture of rural and urban population, while Orlando Health  serves more Hispanics and people living in urban settings. 
Methods to account for heterogeneity across sites are critically needed.

The second major challenge is structural missingness. The prevailing assumption in most of the existing research is that data structures are uniform across sites, meaning each site has the same set of variables available for analysis. However, variations in data collection protocols and resource levels for data linkage lead to discrepancies in the available variable sets at different sites. 
For instance, when using data from Alzheimer’s Disease Research Centers (ADRCs) \citep{NACC} across the United States to investigate the factors associated with Alzheimer’s Disease (AD),
specific data domains are collected only at certain sites, 
resulting in structural missingness on a site-by-site basis. An important example is the complete absence of patients' genetic information, as genetic testing was not included in the original study designs at certain sites. However, omitting genetic data from the process of identifying risk factors for AD 
is problematic, given that genetic variants like the Apolipoprotein E4 (APOE4) allele are recognized as significant contributors to AD development.

In addressing each of these two main challenges, important efforts have been made. To account for between-site distribution heterogeneity, for example,
\cite{cai2021individual} proposed an integrative method for high-dimensional data under sparse regression models that accommodates study-specific differences in covariate distributions and model parameters. \cite{duan2019heterogeneity} proposed a method for heterogeneity-aware 
distributed inference, which permits site-specific nuisance parameters and uses a density ratio tilting technique to handle between-site distribution heterogeneity.  
On the other hand, to tackle the issues caused by structurally missing data, for example, \cite{kundu2019generalized} developed GENMETA, a generalized method of moments approach that combines summary data from multiple studies, allowing for structural missingness but assuming homogeneity in data distributions across studies. \cite{han2023integrating} proposed an empirical-likelihood-based framework to integrate information from multiple studies, while also assuming homogeneity across studies. More methods that have the homogeneity assumption include \cite{chatterjee2016constrained, han2019empirical, gu2019synthetic}, and this assumption can lead to significant biases in the presence of data heterogeneity.

Developing distributed inference methods that simultaneously address both distribution heterogeneity and structural missingness presents several complexities. Firstly, most methods that can manage structural missingness alongside heterogeneity require either the imposition of additional distributional assumptions regarding heterogeneity types or the incorporation of supplementary data to characterize differences across sites. A prevalent assumption is the availability of a reference sample at each site, encompassing all covariates. 
Similar assumptions were made in other data integration research, as seen in the work of \citet{chatterjee2016constrained} and \citet{han2019empirical}. Secondly, unlike the settings considered in \cite{chatterjee2016constrained} and \cite{han2019empirical}, for distributed inference, the reference samples from each site often cannot be combined due to the same regulatory constraints preventing the pooling of patient-level data. The inability to share individual-level data means that many conventional methods relying on pooled data are not directly applicable, necessitating the development of new approaches.

To address the aforementioned challenges, we develop a distributed inference framework tailored for data that is both heterogeneously distributed and structurally missing, utilizing a density-tilted generalized method of moments method. Through reference samples at different sites, we introduce procedures to estimate and communicate the density of the covariates at each site. We devise our procedure to be communication-efficient where aggregated data are exchanged only once. We show that our method performs well in simulation studies, while the existing methods that ignore data heterogeneity can have substantial biases. 

The rest of the paper is organized as follows. Section 2 presents our framework, where we introduce a general setting in distributed inference in the presence of heterogeneity in both covariate distribution and data structure. Our method, the computational algorithm, and asymptotic properties are also introduced in Section 2. In Section 3, we evaluate our method and compare it with some alternative methods via simulation studies. 
Finally, we conclude this paper by a discussion in Section 4.

\section{Model and Method}
\label{sec:Method}

\subsection{Notation and Problem Setup}\label{sec:notation}

Let $\X$ denote a $p$-dimensional covariate 
and $Y$ denote a scalar outcome. 
Assume a parametric model $Y|\X \sim f(y|\x;\bb)$ indexed by a $p$-dimensional parameter $\bb$, is shared across sites. 
We are interested in estimating $\bb$, with its true value denoted as $\bb^*$. 
Suppose there are $K$ independent sites. To allow for structural missingness, at site $j$, we have a study sample $\mathcal{D}_j = \{(\x^{(j)}_{i},y_{ij}), i=1,\dots,N_j\}$, which are independent and identically distributed replicates of $(\X^{(j)},Y)$, where $\X^{(j)} \subseteq \X$ is a subset of $\X$ with dimension $q_j$. In addition, we have a reference sample $\mathcal{D}^*_j = \{\x^*_{ij}, i=1,\dots,n^*_j\}$ with $\x^*_{ij}$s being independent and identically distributed replicates of $\X$ at site $j$. 
This reference sample only contains covariate information, and allows the estimation of the site-specific marginal distribution of $\X$. We denote $q = \sum_{j=1}^K q_j$ and assume $q \geq p$ {and $\cup_{j}\X^{(j)} = \X$}. Our goal is to integrate information from all sites to estimate $\bb^*$ 
without exchanging individual-level  data across sites. 

\subsection{Estimation Procedure}

We start with a procedure similar to that of  \citet{chatterjee2016constrained}.  At site $j$, unless the full covariate $\X$ is observed in the study sample, the regression model $Y|\X \sim f(y|\x;\bb)$ cannot be fit directly. In the presence of structural missingness, we can fit a reduced model $Y |\X^{(j)} \sim h_j(y|\x^{(j)}; \bt_j)$ and get the maximum likelihood estimator of $\bt_j$ as $\widehat{\bt}_j$, where $\bt_j$ is a $q_j$ dimensional parameter. We emphasize here that the reduced model $h_j(\cdot|\cdot)$ does not have to be correctly specified or be coherent  with the main model $f(y|\x;\bb)$, hence the estimator $\widehat{\bt}_j$ only converges to some limiting value $\bt^*_j$, which satisfies $E_{(\X,Y)}\{\partial \log h_j(Y|\X^{(j)};\bt_j)/\partial \bt_j|_{\bt_j=\bt_j^*}\}=\0$ by M-estimation theory.  
We define the estimating function $\s_j(\cdot)$ as
\begin{align}
\label{eq:sj}
\s_j(x;\bb,\bt_j) = \int \left\{\partial \log h_j(y|\x^{(j)};\bt_j)/  \partial \bt_j\right\} f(y|\x;\bb) dy.
\end{align}
Then regardless of whether the reduced model $h_j(y|\x^{(j)};\bt_j)$ is correct or not, $(\bb^*, \bt^*_j)$ satisfies $E_j\{\s_j(X;\bb^*,\bt^*_j)\}$ $=\0$, where $E_j(\cdot)$ denotes the expectation with respect to the true density of $\X$ at site $j$, denoted as $f_j(\cdot)$. It is important to recognize that the information on $\bb^*$ at site $j$ is summarized into $E_j\{\s_j(\X;\bb^*,\bt^*_j)\}=\0$. 
If study samples from all sites were pooled together, the estimating equation $E_j\{\s_j(X;\bb^*,\bt^*_j)\}$ $=\0$ 
for each site can be combined together to estimate the common parameter of interest $\bb$. In the distributed data setting where the individual-level data cannot be shared, we need to derive a distributed inference procedure. 

Without loss of generality, assume site 1 is the lead site where we have access to individual-level data, and only aggregated data from other sites can be shared. To communicate information on $\bb^*$ from sites $j=2,\dots,K$ and adjust for the between-site heterogeneity in $f_j(\cdot)$, we propose the following density ratio tilting method by rewriting $E_j\{\s_j(\X;\bb^*,\bt^*_j)\}=\0$ as
$$E_1\left\{\s_j(\X;\bb^*,\bt^*_j)f_j(\X)/f_1(\X)\right\}=\0, \ \ j=2,\ldots,K.$$
Therefore, at site 1, we have a set of $q$-dimensional estimating functions 
\begin{align}
\label{eq:g}
\g(\X;\bb,\bt^*,\f) = \left\{s_1^{\T}(\X;\bb,\bt^*_1),s_2^{\T}(\X;\bb,\bt^*_2)\frac{f_2(\X)}{f_1(\X)},\dots,s_K^{\T}(\X;\bb,\bt^*_K)\frac{f_K(\X)}{f_1(\X)}\right\}^{\T},
\end{align}
where $\bt^* = (\bt^{*T}_1,\dots,\bt^{*T}_K)^{T}$ and $ \f=\{f_1(\cdot),\dots,f_K(\cdot)\}$. 
Now we have $E_1\{\g(\X;\bb^*,\bt^*,\f)\}=\0$, which can be used to construct the estimation procedure for $\bb$ as outlined below.

Using the reference sample $\mathcal{D}^*_1$ at site 1, we have the $q$-dimensional estimating equation
\begin{equation}
\label{eq:gbar}
\overline{\g}(\bb; \bt^*, f) = (1/n^*_1)\hbox{$\sum_{i=1}^{n^*_1}$}  \g(\x^*_{i1};\bb, \bt^*,\f)=\0.
\end{equation}
Since the dimension of the estimating equation is higher than the dimension of the parameter of interest, i.e., $q \geq p$, we propose to use the  generalized method of moments estimator \citep{hansen1982large} to estimate  $\bb^*$ as 
$$\widehat{\bb}_{\W,1} = \textrm{arg}\min\limits_{\bb} \overline{\g}^{\T}(\bb; \bt^*, \f)  \W \overline{\g}(\bb; \bt^*, \f),$$
where $\W$ is a $q \times q$ positive-semidefinite weight matrix.

Using generalized method of moments theory, we derive the asymptotic  properties of $ \widehat{\bb}_{\W,1}$ and summarize it in the following proposition. 

\begin{proposition}
 Under regularity conditions for model $f(y|\x;\bb)$ and conditions 1-7 in Supplementary Material S1, $n^{*1/2}_1 (\widehat{\bb}_{\W,1} - \bb^*)$ converges in distribution to a normal distribution with mean $\0$ and variance $(\A^{\T} \W \A)^{-1}\A^{\T} \W \bOmega \W^{\T} \A (\A^{\T} \W \A)^{-1}$, where $ \A = E_1\{\partial \g(\X;\bb,\bt^*,\f)/\partial \bb|_{\bb=\bb^*}\}$ and $\bOmega= E_1\{\g(\X;\bb^*,\bt^*,\f) \g^{\T}(\X;\bb^*,\bt^*,\f)\}$. The optimal weight matrix is $\W_{opt,1} = \bOmega^{-1}$, and the corresponding estimator converges in distribution to a normal distribution with mean $\0$ and covariance matrix $(\A^{\T} \bOmega^{-1} \A)^{-1}$.
\end{proposition}

Note that, till now we assume that $\bt^*_j$ and $f_j(\cdot)$ are known, which is not true in practice. Next we discuss the estimation of $\bt^*_j$ and $f_j(\cdot)$ under the distributed data setting, and their impact on the estimation of $\bb^*$.

\subsection{Estimation of $\bt^*_j$}
In this sebsection, we assume $f_j(\cdot)$ is known and consider the estimation of $\bt_j^*$. At site $j$, $\bt^*_j$ can be consistently estimated by $\widehat{\bt}_j$, which is the solution to the estimating equation 
\begin{align}
\label{eq:reduced}
\hbox{$\sum_{i=1}^{N_j}$} {\partial \log h_j(y_{ij}|\x_i^{(j)};\bt_j)/ \partial \bt_j }= \0.
\end{align}
Under regularity conditions for $h_j(y|\x^{(j)};\bt_j)$, $N^{1/2}_j (\widehat{\bt}_j - \bt^*_j) \to N(\0,\bSigma_j)$ in distribution. Here $\bSigma_j$ is the asymptotic covariance matrix and can be estimated using the {robust sandwich variance estimator} $\widehat{\bSigma }_j$. 

With $\bt^*$ estimated by $\widehat{\bt}$, we can construct the $q$-dimensional estimating equation $\overline{\g}(\bb;\widehat{\bt}, \f)$ using reference sample $\mathcal{D}^*_1$ at site 1 and estimators $\widehat{\bt}_j, j=1,\dots,K$. The corresponding generalized method of moments estimator of $\bb^*$ is defined as 
$\widehat{\bb}_{\W,2} = \textrm{arg}\min\limits_{\bb} \overline{\g}^{\T}(\bb;\widehat{\bt}, \f) \W \overline{\g}(\bb;\widehat{\bt}, \f),
$
where $\W$ is a $q \times q$ positive-semidefinite weight matrix. We now summarize the asymptotic  properties of $\widehat{\bb}_{\W,2}$ in the following proposition.  
\begin{proposition}
Denote $\B_j= E_j\{ {\partial \s_j(\X;\bb^*,\bt_j)/ \partial \bt_j }|_{\bt_j=\bt_j^*} \}$, $c_{j} = \lim\limits_{n^*_{1} \to \infty} n^*_{1}/N_j$, and $\bGamma_j = c_j \B_j \bSigma_j \B^{\T}_j$ with  $\bGamma=\mbox{diag}(\bGamma_1,\ldots ,\bGamma_K)$.
 Under regularity conditions 1-9 in Supplementary Material S1, $n^{*1/2}_{1} (\widehat{\bb}_{\W,2}-\bb^*)$ converges in distribution to the normal
distribution with mean $\0$ and covariance matrix $$(\A^{\T} \W \A)^{-1}\A^{\T} \W (\bOmega+\bGamma) \W^{\T} \A (\A^{\T} \W \A)^{-1}.$$ The optimal weight matrix is $\W_{opt,2} = (\bOmega+\bGamma)^{-1}$, and the corresponding estimator converges in distribution to the normal distribution with mean $\0$ and covariance matrix $\{\A^{\T} (\bOmega+\bGamma)^{-1} \A\}^{-1}$.
\end{proposition}

As in \citet{hansen1996finite}, we iteratively update the unknown parameters in the optimal weight matrices $\W_{opt,1}, \W_{opt,2}$ and solve the optimization problem to obtain  $\widehat{\bb}_{\W,1}, \widehat{\bb}_{\W,2}$.

\subsection{Estimation of $f_j(\cdot)$}
Finally we consider the estimation of $\{f_j(\cdot), j =1,\ldots, K\}$, i.e., the density of $\X$ at each site. As defined in Section \ref{sec:notation}, the reference sample $\mathcal{D}^*_j$ at site $j$ is a random sample of $\X$ at site $j$. For the lead site, $f_1(\cdot)$ can be estimated using any standard  nonparametric density estimator, denoted as $\widehat{f}_1(\cdot)$, based on the reference sample $\mathcal{D}^*_1$. For external sites $j=2,\dots,K$, since individual-level data of the reference sample are not accessible to site 1,  the nonparametric density estimators for $\{f_j(\cdot), j=2,\dots,K\}$ cannot be constructed directly at site 1. 
Moreover, although these multivariate density estimators can be constructed using reference samples at external sites, they could be expensive to communicate to site 1 due to the curse of dimensionality. For instance, to communicate a $p$ dimensional density function at the grid points formed by a mesh with size $m_1\times \dots\times m_p$ {with $m_i$ being the number of pre-specified points in the support of the $i$-th covariate}, we will need to communicate $\prod_{i=1}^p m_i$ values of the density function.

We adopt a copula density estimation method to overcome these difficulties \citep{yan2007enjoy}.
The copula approach reduces the multidimensional estimation problem to a one-dimensional problem and simultaneously eases the aforementioned difficulty in data transmission. {In the above  example, to communicate the density values at the same grid points, we will only need to transport $\sum_{i=1}^pm_i$ values.}
Specifically, denote $\X=(X_1,\dots,X_p)$ where $X_i$ is the $i$-th covariate, and denote the cumulative distribution of $\X$ at site $j$ as $F_j(\cdot)\equiv F_j(x_1,\dots,x_p)$ with marginal distribution functions $F_{j1}(x_1),\dots,F_{jp}(x_p)$.  Assume that
$F_j(x_1,\dots,x_p) = C\{F_{j1}(x_1),\dots,F_{jp}(x_p);\alpha_j\},$
where $C$ is the copula and $\alpha_j$ is the correlation parameter. Then $F_j(\cdot)$ can be estimated as
\begin{align}
\label{eq:coupla}
\widehat{F}_j(x_1,\dots,x_p) &= C\{\widehat{F}_{j1}(x_1),\dots,\widehat{F}_{jp}(x_p), \widehat{\alpha}_j\},
\end{align}  
where $\widehat{F}_{j1}(x_1),\dots,\widehat{F}_{jp}(x_p)$ are the empirical marginal distributions and $\widehat{\alpha}_j$ is the maximum likelihood estimator after plugging in $\widehat{F}_{j1}(x_1),\dots,\widehat{F}_{jp}(x_p)$ using reference sample $\mathcal{D}^*_j$ at site $j$. {Although  $\widehat{F}_j(x_1,\dots,x_p)$ can be constructed at site $j$, 
to reduce the communication cost, we only estimate  $\widehat{F}_{j1}(x_1), \dots, \widehat{F}_{jp}(x_p), \widehat\alpha_j$ at site $j$ and transport them to site 1, and then construct 
$\widehat{F}_j(x_1,\dots,x_p)$ at site 1. We denote the corresponding density estimator as $\widehat{f}_j(\cdot)$.}
The choice of $C$ and the correlation parameter $\alpha_j$ can be flexible. For example, one can use the
fully nested Archimedean copula proposed in \cite{hofert2011nested}. 

With $\bt^*$ estimated by $\widehat{\bt}$ and $f_j(\cdot)$ estimated by $\widehat{f}_j(\cdot)$ {using 
the  copula-based method,} we can construct the $q$-dimensional estimating equation $\overline{\g}(\bb;\widehat{\bt}, \widehat{\f})$. The corresponding generalized method of moments estimator of $\bb^*$ is defined as 
\begin{align}
\label{eq:estimator}
\widehat{\bb}_{\W} = \text{arg}\min\limits_{\bb} \bar{\g}^{\T}(\bb;\widehat{\bt},\widehat{\f}) W \overline{\g}(\bb;\widehat{\bt}, \widehat{\f}),
\end{align}
where $\W$ is a $q \times q$ positive-semidefinite weight matrix and $\widehat{\f}=\{\widehat{f}_1(\cdot),\widehat{f}_2(\cdot),\dots,\widehat{f}_K(\cdot)\}$. 
We summarize the details into Algorithm 1 in the following subsection, and report the asymptotic  properties of $ \widehat{\bb}_{\W}$ in Theorem 1. 

\begin{theorem}\label{thm1}
Under regularity conditions in Supplementary Material, 
$n^{*1/2}_{1} (\widehat{\bb}_{\W}-\bb^*)$ converges in distribution to the normal distribution with mean $\0$ and covariance matrix $$(\A^{\T} \W \A)^{-1}\A^{\T} \W (\bOmega+\bGamma+\V) \W^{\T} \A (\A^{\T} \W \A)^{-1},$$ where $\V$ is given in Supplementary Material. The optimal weight matrix is $\W_{opt,3} = (\bOmega+\bGamma+\V)^{-1}$, and the corresponding estimator converges in distribution to the normal distribution with mean $\0$ and covariance matrix $\{\A^{\T} (\bOmega+\bGamma+\V)^{-1} \A\}^{-1}$.
\end{theorem}

\subsection{Our Algorithm}
To present the details of our algorithm, at site $j$, we categorize the aggregated 
information into two categories:
\begin{enumerate}
\item Estimated parameters. Denote by  $\mathcal{P}_j=(\hat{\bt}_j,\hat{\bSigma}_j)$ as the estimated parameters of the reduced model $h_j(y|\x^{(j)}; \bt_j)$ and the relevant robust sandwich estimator. 
\item Density of $\X$. Denote by $\mathcal{G}_k = \{x^1_k, \dots,x^{m_k}_k\}$ a set of $m_k$ pre-specified points in the support of each covariate $X_k, k=1,\dots,p$, and $\mathcal{G} = \{\mathcal{G}_1,\dots,\mathcal{G}_p\}$. Practically, these points can be chosen as regularly spaced grid points in the support of each covariate or based on quantiles of the distribution of each covariate. Denote by $\{\widehat{F}_{j1}(x_1), \dots, \widehat{F}_{jp}(x_p)\}$ the empirical marginal distribution of $\X$ using reference sample $\mathcal{D}^*_j$ at site $j$. Denote by $\hat{F}_j(\mathcal{G})= \{\hat{F}_{j1}(\mathcal{G}_1), \dots, \hat{F}_{jp}(\mathcal{G}_p)\}$ and $\hat{F}_j =\{\hat{F}_j(\mathcal{G}), \hat{\alpha}_j, \mathcal{G}\}$ the aggregated information about $F_j(\cdot)$ using the copula. 
\end{enumerate}

We summarize the estimation procedure in Algorithm 1. 

\begin{algorithm}
\caption{The 
{{distributed}} density-tilted {{generalized method of moments}} estimator.}
\begin{algorithmic}[1]
\FOR{$j$ in \{$2,...,K$\}}
\STATE Fit the reduced model using  study sample $\mathcal{D}_j$ to obtain $\mathcal{P}_j$ according to {{equation}} (\ref{eq:reduced}).
\STATE Obtain $\hat{F}_j(\mathcal{G})$ and $\hat{\alpha}_j$ using reference sample $\mathcal{D}^*_j$.
\STATE Transfer $\{\mathcal{P}_j, h_j(\cdot),
\hat{F}_j\}$ to site 1.
\ENDFOR
\STATE For Site 1, 
\STATE \quad Obtain $\mathcal{P}_1$ using study sample $\mathcal{D}_1$  according to  {{equation}} (\ref{eq:reduced}) and $\hat{f}_1(\cdot)$ using $\mathcal{D}^*_1$.
\STATE \quad Obtain $\widehat{F}_{j1}(x_1),\dots,\widehat{F}_{jp}(x_p)$ using $\{\widehat{F}_{j1}(\mathcal{G}_1), \dots, \widehat{F}_{jp}(\mathcal{G}_p)\}$, for $j=2,\dots,K$.
\STATE \quad Obtain $\widehat{F}_j(\cdot)$ according to equation~(\ref{eq:coupla}) and the corresponding density estimator $\hat{f}_j(\cdot)$. 
\STATE \quad Obtain $\hat{\bb}_W$ and its estimated covariance matrix  according to equations~(\ref{eq:sj}),
    (\ref{eq:gbar}) and (\ref{eq:estimator}).
\STATE Output: $\hat{\bb}_W$ and its estimated covariance matrix.
\end{algorithmic}
\end{algorithm}

\begin{remark}\label{remark1} 
{An alternative to estimate $f_j(\cdot)$ is via a synthetic data approach.} Using reference sample $\mathcal{D}^*_j$ at site $j$, we can estimate the density of $\X$ nonparametrically and simulate synthetic data from the estimated density. Individual-level synthetic data are then transferred to site 1 and the density ratios can be calculated based on the empirical distribution of the synthetic data. The advantage of the synthetic data approach is that it is easy to implement and it does not require any assumptions on the distribution of $\X$. {
A potential limitation of this approach is that multivariate density estimator requires a large sample size to work well. When reference sample is small, the estimates of $\bb$ could be biased. Moreover, generating synthetic data could also inflate the variance of the estimates of $\bb$ by introducing extra sampling variation. These issues will be investigated in our simulation study.} 
\end{remark}

\begin{remark}\label{remark2}
Our method differs from \citet{chatterjee2016constrained} and \citet{han2019empirical} in that we consider a more general setting by targeting joint learning of a global full model across multiple studies where structural missingness may exist in all studies. 
In contrast, \citet{chatterjee2016constrained} and \citet{han2019empirical} focus on leveraging information from a single external study to improve learning efficiency in the internal study that has complete data, making our method be more general and flexible. 
More importantly, by utilizing a copula approach, our method eliminates the need to share individual-level reference data for estimating covariate densities, while the methods proposed by \citet{chatterjee2016constrained} and \citet{han2019empirical} require such data sharing. This distinction enhances the applicability of our method in real-world research networks, where individual-level data sharing is often restricted.
\end{remark}

\section{SIMULATIONS}
We evaluate the performance of our method through simulation studies. 
{For illustration, we consider a scenario with $K=3$ sites.} Assume that the relationship between a binary outcome variable $Y$ and three covariates $\X=(X_1, X_2, X_3)^{\T}$ is described by the logistic regression model
$\hbox{logit}\{\pr(Y|\X)\}=\beta_0+\beta_1 X_1 + \beta_2 X_2 + \beta_3 X_3,$
where $X_1$ is a binary covariate, and $X_2,X_3$ are continuous covariates. We set $\beta_0=-3,\beta_1=1,\beta_2=1,\beta_3=1$. 

Given the binary covariate $X_1$, $(X_2,X_3)$ are generated from the Clayton copula family with marginal beta distributions. The baseline setting we considered here assumes a homogeneous population across sites. In this setting, the marginal distribution of $X_1$, $X_2$ and $X_3$, as well as the correlation among them, remains the same across three sites. To illustrate distribution heterogeneity, we also make the distribution of the covariates to be different across sites. Specifically, we have two partially heterogeneous settings, with either the marginal distribution of $X_{1}$ (partially heterogeneous setting 1) or the joint distribution of $X_{2}$ and $X_{3}$ for each level of $X_{1}$ (partially heterogeneous setting 2) to be different across sites, and a completely heterogeneous setting with both parts being different across sites. The details can be found in the Supplementary Material S2. To illustrate, we show the contour plots of $(X_2,X_3)$ under different levels of $X_1$ at each site under the completely heterogeneous setting in Figure \ref{fig1}.  

\begin{figure}
	\centering
	\includegraphics[width = 0.8\linewidth]{contour2.pdf}
	\caption{\baselineskip=12pt Joint distribution of $(X_2,X_3)$ at each site given $X_1=1$ (top panels) and $X_1=\0 $ (bottom panels). Probabilities of $X_1=1$ at site 1, 2 and 3 are $0.2,0.5,0.8$, respectively. 
	}
\label{fig1}
\end{figure} 
 
To generate structurally missing data, assume at site 1, we observe $(X_{1},X_{2},Y)$ with sample size $N_1$;  at site 2, we observe $(X_{1},X_{3},Y)$ with sample size $N_2$;  at site 3, we observe $(X_{2},X_{3},Y)$ with sample size $N_3$ and we let $N_1=N_2=N_3=1000$. In addition, at sites $j=1,2,3$, we observe reference samples $\mathcal{D}^*_j = \{\x^*_{ij}=(x^*_{ij1}, x^*_{ij2}, x^*_{ij3})^T, i=1,\dots,n^*_j\}$ with $n^*_1=n^*_2=n^*_3=n$ whose value changes with the setting.  
In order to provide a comprehensive guidance for the implementation of our method, we experiment with
\begin{itemize}
    \item[I.] a set of reference sample size $n$ to investigate its impacts on density estimation; 
    \item[II.] multiple values of the grid density $m_i$ to investigate its impacts on the performance of our method;
    \item[III.] several different methods to evaluate the robustness of each method to the structural missingness and distributional shifts. The methods compared include
\begin{itemize}
    \item[1.] Our copula-based method, abbreviated as dist-GMM-C;
    \item[2.] The method proposed by \cite{kundu2019generalized}, which handles structural missingness but assumes the data are homogeneously distributed, abbreviated as GENMETA;
    \item[3.] The synthetic data-based density-tilted generalized method of moments approach mentioned in Remark \ref{remark1}, abbreviated as dist-GMM-S;
    \item[4.] Local analysis using the data at site 1 only, abbreviated as Local.
\end{itemize}
\end{itemize}
For our dist-GMM-C method, the densities are estimated by the Clayton copula model: we used the R package copula \citep{copula} for copula density estimation. We repeated the simulation $100$ times for each setting.

First, to find a range of the reference sample size $n$ for ensuring satisfactory performance of density estimation, we varied the value of $n$ from 50 to 1,000 and assessed the root mean squared error (RMSE) for estimating the coefficient parameter vector $(\beta_1,\beta_2,\beta_3)^T$. 
From Figure S1 in Supplementary Material S3, we observed a sharp decrease in RMSE for dist-GMM-C when $n$ increased from 50 to 100. When $n$ is larger than 300, the RMSE for dist-GMM-C did not exhibit much further reduction. Notably, 300 is less than one third of the study sample size within each site, reflecting a limited need for a very large reference set. On the other hand, as expected, 
dist-GMM-S requires more reference samples to achieve accurate estimation.
To guarantee a good performance, 
we fix $n$ at 500 for the following simulations.

We then investigate the effects of grid density $m_i$ on dist-GMM-C. We explore three different values of $m_i$: 50, 100, and 200, and examine their impacts on the estimation and inference results for each coefficient. The results are reported in Table S1 of Supplementary Material S3. It turns out that when $m_i$ was set to 100, dist-GMM-C yielded estimates with small biases and the coverage rate of the 95\% confidence interval approached 95\% for most cases. Overly increasing $m_i$ can lead to inflated biases and variances, and it also increases the communication cost. 
Therefore, we fix $m_i$ at 100 for further comparisons between methods.

\begin{figure}
	\centering
	\includegraphics[width = 0.8\linewidth]{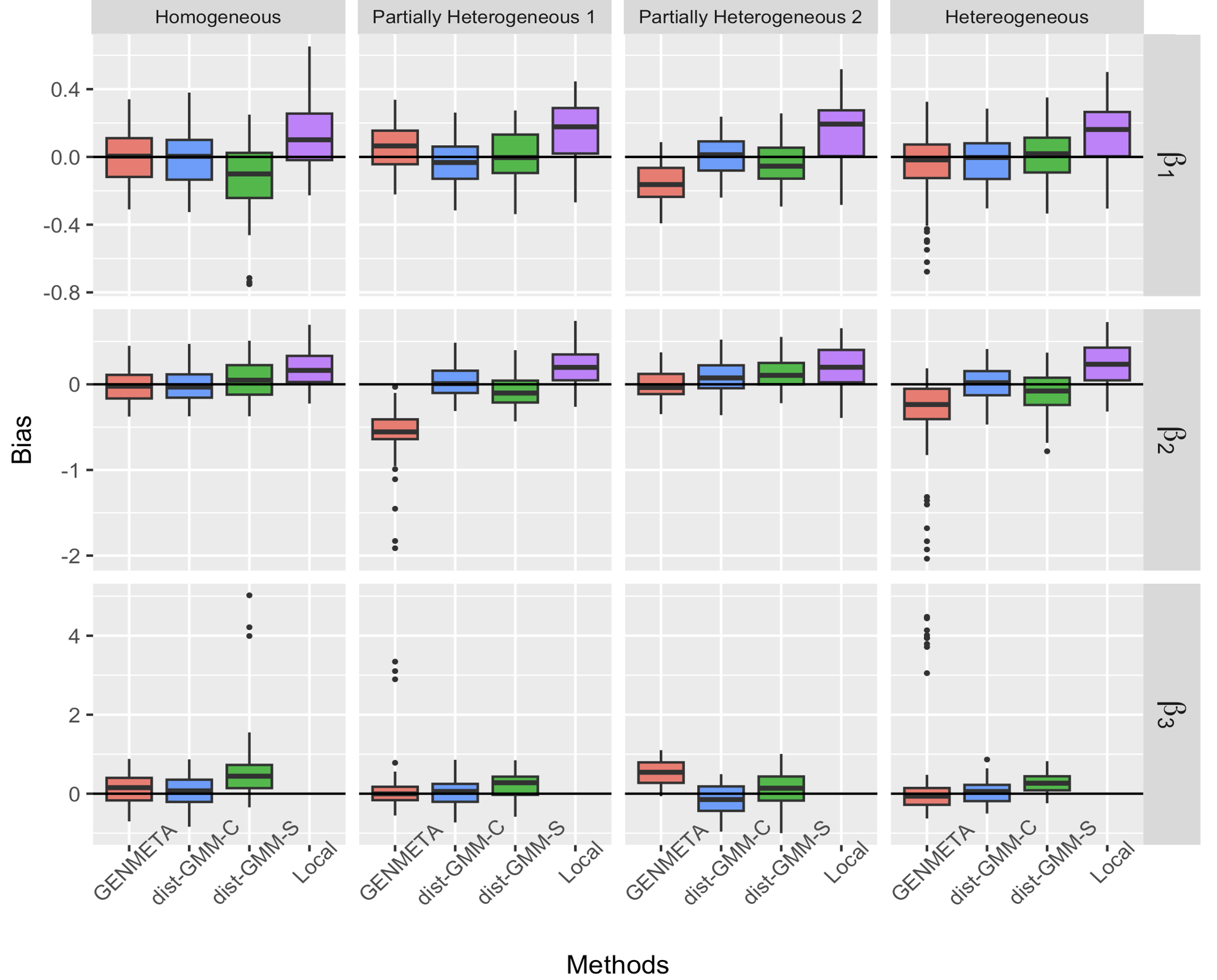}
	\caption{\baselineskip=12pt Boxplots of the empirical biases of four methods for estimating parameters $\beta_1$ (top), $\beta_2$ (middle) and $\beta_3$ (bottom) under four simulation settings.
	}
\label{simu-fig3}

\end{figure}

Finally, with $n=500$ and $m_i=100$, we conduct a comparison of the estimation and inference performance of all four methods. The empirical biases of the estimates for each of three coefficients are presented in Figure \ref{simu-fig3}, encompassing all four simulation settings. The summary statistics as well as the coverage probabilities of the 95\% confidence intervals are shown in Table \ref{tab:comparison}. 
Under the homogeneous setting, both dist-GMM-C and GENMETA exhibit small biases, small variances, and good coverage rates. It is noted that, dist-GMM-C shows slightly larger variability than GENMETA, suggesting some inflation of variability due to density ratio estimation when there is no actual distribution shift present. 
In cases where a distribution shift exists, the biases of dist-GMM-C remain small; while the biases of GENMETA increase and the coverage rates of its confidence intervals deteriorate. 
When comparing dist-GMM-S to dist-GMM-C,  since the former introduces both the bias in multivariate density estimation and the extra sampling variation from generating synthetic data into parameter estimation, it exhibited noticeable biases and higher variability in most cases, rendering its estimates less reliable. 
Finally, the local analysis results generally exhibit significant biases and larger variability compared to dist-GMM-C in estimation, and the coverage rates of the corresponding confidence intervals are lower than the desired level of 0.95. In addition, it cannot provide any results for the missing covariate ($\beta_3$ in our simulation). 

\begin{table}[hbt!]
\centering
\caption{\baselineskip=12pt Summary of method comparison: estimated biases (Bias), standard errors (SD), estimated standard errors (ESD), and coverage probabilities (CI) of 95\% confidence intervals of parameter estimates. Setting 1 represents the homogeneous setting, setting 2 represents the partially heterogeneous setting 1, setting 3 represents the partially heterogeneous setting 2, and setting 4 represents the heterogeneous setting. \label{tab:comparison}}
\resizebox{\textwidth}{!}{\begin{tabular}{crrrrrrrrrrrrrrrrr}
  \hline
  &  & \multicolumn{4}{c}{GENMETA} & \multicolumn{4}{c}{dist-GMM-C} & \multicolumn{4}{c}{dist-GMM-S} & \multicolumn{4}{c}{Local} \\
   \cline{3-6}   \cline{7-10}  \cline{11-14}  \cline{15-18}
  Setting & $\bb$ & Bias & SD & ESD & CI & Bias & SD & ESD & CI & Bias & SD & ESD & CI & Bias & SD & ESD & CI \\
  \hline
& $\beta_1$ & 0.00 & 0.16 & 0.16 & 0.95 & 0.00 & 0.16 & 0.16 & 0.95 & -0.12 & 0.20 & 0.16 & 0.90 & 0.12 & 0.19 & 0.20 & 0.90 \\
1 &  $\beta_2$ & -0.02 & 0.18 & 0.23 & 0.99 & -0.01 & 0.19 & 0.25 & 0.99 & 0.05 & 0.22 & 0.26 & 0.98 & 0.19 & 0.23 & 0.27 & 0.89 \\ 
&  $\beta_3$ & 0.13 & 0.36 & 0.42 & 0.96 & 0.08 & 0.41 & 0.38 & 0.92 & 0.54 & 0.83 & 0.55 & 0.92 &  &  &  &  \\ 
& $\beta_1$ & 0.05 & 0.13 & 0.14 & 0.98 & -0.03 & 0.14 & 0.19 & 1.00 & 0.01 & 0.15 & 0.19 & 1.00 & 0.15 & 0.19 & 0.19 & 0.89 \\ 
2 & $\beta_2$ & -0.56 & 0.29 & 0.21 & 0.27 & 0.03 & 0.18 & 0.20 & 0.98 & -0.09 & 0.19 & 0.22 & 1.00 & 0.22 & 0.24 & 0.27 & 0.85 \\ 
& $\beta_3$ & 0.08 & 0.62 & 0.30 & 0.98 & 0.04 & 0.38 & 0.47 & 1.00 & 0.21 & 0.32 & 0.46 & 1.00 &  &  &  &  \\ 
& $\beta_1$ & -0.15 & 0.12 & 0.14 & 0.84 & 0.00 & 0.12 & 0.14 & 1.00 & -0.04 & 0.13 & 0.15 & 1.00 & 0.15 & 0.19 & 0.21 & 0.91 \\ 
3 & $\beta_2$ & -0.01 & 0.17 & 0.22 & 1.00 & 0.07 & 0.20 & 0.20 & 0.93 & 0.12 & 0.18 & 0.22 & 0.96 & 0.20 & 0.26 & 0.28 & 0.92 \\ 
& $\beta_3$ & 0.52 & 0.31 & 0.41 & 0.79 & -0.15 & 0.38 & 0.40 & 0.96 & 0.14 & 0.41 & 0.58 & 1.00 &  &  &  &  \\ 
& $\beta_1$ & -0.05 & 0.19 & 0.13 & 0.94 & -0.02 & 0.14 & 0.17 & 1.00 & 0.01 & 0.14 & 0.18 & 0.98 & 0.14 & 0.19 & 0.21 & 0.89 \\ 
4 & $\beta_2$ & -0.33 & 0.45 & 0.24 & 0.93 & 0.02 & 0.21 & 0.25 & 1.00 & -0.09 & 0.25 & 0.28 & 0.97 & 0.22 & 0.24 & 0.28 & 0.91 \\ 
& $\beta_3$ & 0.29 & 1.24 & 0.28 & 0.98 & 0.04 & 0.29 & 0.43 & 1.00 & 0.26 & 0.26 & 0.45 & 1.00 &  &  &  &  \\ 
   \hline
\end{tabular}}
\end{table}

In summary, our simulation study has demonstrated the superior performance of our dist-GMM-C method when integrating data with heterogeneity and structural missingness, offering several practical implications. First, among all the methods considered, GENMETA, as expected, shows sizable bias in the presence of distribution heterogeneity; the local analysis yields biased estimates with considerable variability and fails to account for missing covariates; similarly, dist-GMM-S presents biased estimates and high variability. In contrast, dist-GMM-C is a robust approach, offering least biased estimates with small variability. Secondly, the application of dist-GMM-C reveals that neither a very large reference dataset nor high grid densities are prerequisites for effective performance. This observation underlines the practicality of our method, suggesting its flexibility and adaptability across different real-world scenarios.

\section{DISCUSSION}

We have introduced a novel framework for distributed inference tailored to distributed data with between-site heterogeneity and structural missingness. Different from existing methods \citep{chatterjee2016constrained, han2019empirical}, our methodology presents a new setting without the necessity of sharing patient-level reference data to address covariate distribution shifts. Utilizing a density-tilted generalized method of moments, our method facilitates the integration of information across multiple sites, accounting for between-site heterogeneity through locally estimated covariate densities using reference samples. Designed for communication efficiency, our algorithm requires only a single exchange of aggregated data. Simulation studies have validated our method's superior robustness to between-site heterogeneity compared to traditional methods, which often yield biased estimates due to their disregard for data heterogeneity. 

A key strategy for addressing covariate distribution heterogeneity across sites is through the density ratio model, which crucially hinges on the availability of reference samples. The availability of such reference samples is becoming increasingly feasible. In multicenter studies that utilize EHR, reference samples can be identified from a subset of patients with more comprehensive data. For example, the OneFlorida network dataset \citep{shenkman2018oneflorida}, focusing on patients at risk for psoriatic arthritis, encompasses around 5,000 patients, with 30\% of their EHRs linked to Medicaid/Medicare claims data. This linkage enriches the dataset with additional information on medication usage, qualifying these patients as a reliable reference sample. Furthermore, for multicenter epidemiological studies, publicly accessible health-related survey data from relevant geographical areas can serve as a reliable source of reference samples \citep{kundu2019generalized}, enhancing the robustness and relevance of the analysis. Additionally, the All of Us initiative \citep{ramirez2022all}, which has enrolled over half a million participants and made their EHR data accessible, with more than 40\% having linked survey and physical activity data, presents a significant resource for enhancing data analysis and interpretation in such studies.

In summary, we have introduced a novel framework for distributed inference that addresses the challenges posed by data heterogeneity 
and missing information across different study sites, without the necessity to share detailed patient-level data. Our approach utilizes a density-tilted generalized method of moments to integrate data from multiple sources, addressing the issue of heterogeneity in data distribution effectively. 
Additionally, the growing availability of extensive health datasets, such as the All of Us initiative, enhances the practicality of applying our methodology to leverage large-scale health data for research. By improving the accuracy and efficiency of distributed inference in the face of heterogeneous datasets, our contribution facilitates more precise identification of disease risk factors, advancing more reliable clinical evidence generation using real-world datasets.







\bibliographystyle{apalike}
\bibliography{ref}

\newpage
\appendix

\centerline{\Large\bf Supplementary Materials}
\vspace{0.3in}

\setcounter{section}{0}
\setcounter{theorem}{0}
\setcounter{figure}{0}
\setcounter{table}{0}

\renewcommand{\thesection}{S\arabic{section}}
\renewcommand{\thetheorem}{S\arabic{theorem}}
\renewcommand{\thelemma}{S\arabic{lemma}}
\renewcommand{\thefigure}{S\arabic{figure}}
\renewcommand{\thetable}{S\arabic{table}}

\theoremstyle{plain}
\newtheorem{theoremS}{Theorem}[section]
\newtheorem{lemmaS}[theoremS]{Lemma}
\newtheorem{propositionS}[theoremS]{Proposition}

\theoremstyle{remark}
\newtheorem{remarkS}[theoremS]{Remark}


\section{Required Regularity Conditions}
We need the following regularity conditions for proposition 1 and proposition 2.
\begin{itemize}
    \item[(1)] The true parameter $\bb^*$ is an interior point of the parameter space $\Theta_{\bb}$ which is compact, and define $\mathcal{N}(\bb^*)$ as a neighborhood of $\bb^*$. 
    \item[(2)] The weight matrix $\W$ is positive semi-definite and satisfies $\W E_1(\g(\X;\bb,\bt^*,\f))=\0$ if and only if $\bb=\bb^*$. 
    \item[(3)] $\s_j(\X;\bb,\bt_j)f_j(\X)/f_1(\X)$ is continuous for each $(\bb,\bt_j)\in \Theta_{\bb} \times \mathcal{N}(\bt_j^*)$ with probability 1, where $\mathcal{N}(\bt_j^*)$ is a neighborhood of $\bt_j^*$ for $j=1,\ldots,K$. 
    \item[(4)] $E_1(sup_{(\bb,\bt_j)\in \Theta_{\bb} \times \mathcal{N}(\bt_j^*)} \| \s_j(\X;\bb,\bt_j)f_j(\X)/f_1(\X) \|) < \infty$ for $j=1,\ldots,K$. 
    \item[(5)] $f_j(\X)/f_1(\X)\partial \s_j(\X;\bb,\bt_j)/ \partial \bb$ is continuous at each $(\bb,\bt_j)\in \mathcal{N}(\bb^*) \times \mathcal{N}(\bt_j^*)$ with probability 1. 
    \item[(6)] $E_1(sup_{(\bb,\bt_j)\in \mathcal{N}(\bb^*) \times \mathcal{N}(\bt_j^*)} \| \partial f_j(\X)/f_1(\X)\s_j(\X;\bb,\bt_j)/ \partial \bb \|) < \infty$. 
    \item[(7)] $\bOmega$ exists and is finite, and $\A$ is of full column rank. 
    \item[(8)] $f_j(\X)/f_1(\X)\partial \s_j(\X;\bb^*,\bt_j)/ \partial \bt_j$ is continuous at each $\bt_j \in \mathcal{N}(\bt_j^*)$ with probability 1. 
    \item[(9)] $E_1(sup_{\bt_j\in \mathcal{N}(\bt_j^*)} \|f_j(\X)/f_1(\X) \partial \s_j(\X;\bb^*,\bt_j)/ \partial \bt_j \|) < \infty$. 
\end{itemize}



\section{Simulation Details}
In order to see the effects of different levels of between-site heterogeneity on the performance of the considered methods, 
we allow the marginal distribution of $X_{1}$ and the joint distribution of $X_{2}$ and $X_{3}$ for each level of $X_{1}$ to be different across sites. Specifically, we have the following four settings:

\begin{itemize}
    \item[1.] Homogeneous setting: across all sites, $X_{1}$ is generated from a Bernoulli distribution with $P(X_1 = 1) = 0.3$. For the marginal distribution of $X_2$ and $X_3$ at each level of $X_1$, we let $X_2 \sim \hbox{Beta}(0.5, 0.5)$ and $X_3 \sim \hbox{Beta}(5, 2)$ for subjects with $X_1=1$, and we let $X_2\sim \hbox{Beta}(0.2, 0.8)$ and $X_3 \sim \hbox{Beta}(2, 2)$ for subjects with $X_1=0$ for all sites. For the correlation strength between $X_2$ and $X_3$, we let $\alpha=1$ across all sites. 
    \item[2.] Partially heterogeneous setting 1: for the marginal distribution of $X_{1}$, we let $P(X_1=1)=0.2$ at site 1, $P(X_1=1)=0.5$ at site 2, and $P(X_1=1)=0.8$ at site 3. There is no heterogeneity in the joint distribution of $X_2$ and $X_3$ at each level of $X_1$ across sites.
    \item[3.] Partially heterogeneous setting 2:  across all sites, we have $P(X_1 = 1) = 0.3$. For the distribution of $X_2$ and $X_3$ at each level of $X_1$, we let $X_2 \sim \hbox{Beta}(0.5, 0.5)$ and $X_3 \sim \hbox{Beta}(5, 2)$ for subjects with $X_1=1$, and we let $X_2\sim \hbox{Beta}(0.2, 0.8)$ and $X_3 \sim \hbox{Beta}(2, 2)$ for subjects with $X_1=0$ for site 1; 
    we let $X_2 \sim \hbox{Beta}(2, 2)$ and $X_3 \sim \hbox{Beta}(1, 2)$ for subjects with $X_1=1$, and let $X_2\sim \hbox{Beta}(1, 2)$ and $X_3\sim \hbox{Beta}(0.2, 0.8)$ for subjects with $X_1=0$ at site 2; 
    and we let $X_2 \sim \hbox{Beta}(2, 5)$ and $X_3 \sim \hbox{Beta}(0.5, 0.5)$ for subjects with $X_1=1$, and we let $X_2 \sim \hbox{Beta}(5, 2)$ and $X_3 \sim \hbox{Beta}(2, 5)$ for subjects with $X_1=0$ at site 3. 
	For the correlation strength between $X_2$ and $X_3$, we let $\alpha_1=1$ at site 1, $\alpha_2=2$ at site 2, and $\alpha_3=3$ at site 3.
 \item[4.] Heterogeneous setting: there exists distribution shift in both the marginal distribution of $X_1$, whose generating follows the same way as in Setting 2, and the joint distribution of $X_2$ and $X_3$ within each level of $X_1$, whose generating follows the same way as in Setting 3.   
\end{itemize}

\clearpage

\section{Additional simulation results}

\begin{figure}
	\centering
	\includegraphics[width = 1\linewidth]{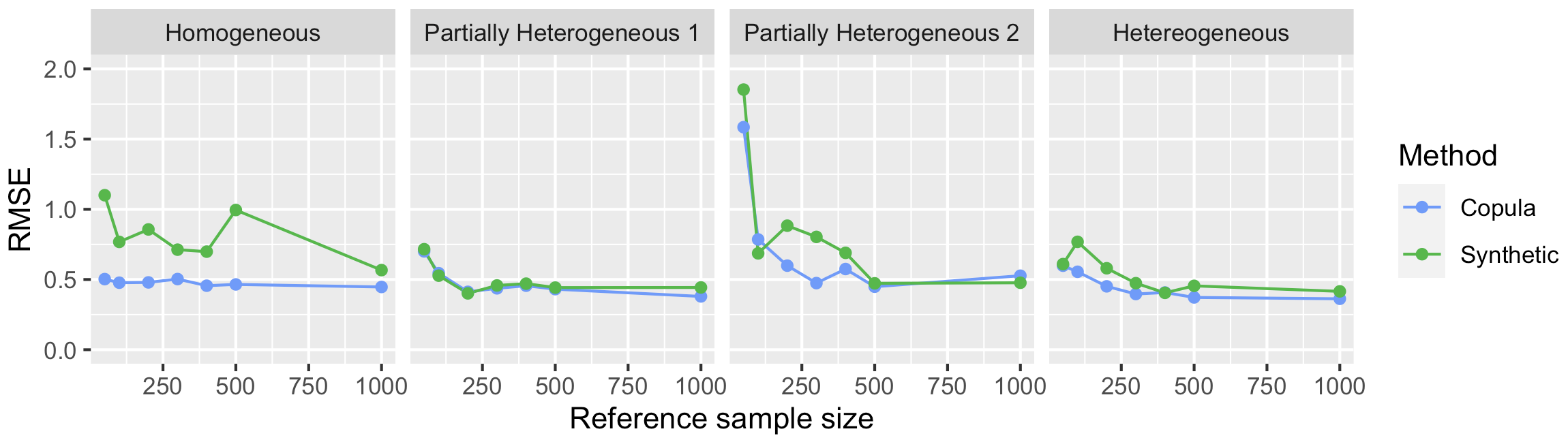}
	\caption{\baselineskip=12pt Line plots of RMSE of dist-GMM-C (Copula) and dist-GMM-S (Synthetic) 
 for estimating $(\beta_1, \beta_2, \beta_3)^T$
 under four simulation settings. The reference sample size $n$ takes value from $\{ 50, 100, 200, 300, 400, 500, 1000 \}$.  
	}
\label{simu-fig1}
\end{figure}

\begin{table}[hbt!]
\centering
\caption{\baselineskip=12pt Summary of grid density $m_i$ selection for dist-GMM-C: estimated biases (Bias), standard errors (SD), estimated standard errors (ESD), and coverage probabilities (CI) of 95\% confidence intervals of parameter estimates. Setting 1 represents the homogeneous setting, setting 2 represents the partially heterogeneous setting 1, setting 3 represents the partially heterogeneous setting 2, and setting 4 represents the heterogeneous setting.  \label{tab:mi}}
\resizebox{\textwidth}{!}{\begin{tabular}{crrrrrrrrrrrrr}
  \hline
  &  & \multicolumn{4}{c}{ $m_i=50$ } & \multicolumn{4}{c}{$m_i=100$} & \multicolumn{4}{c}{$m_i=200$}  \\
   \cline{3-6}   \cline{7-10}  \cline{11-14}  
  Setting & $\bb$ & Bias & SD & ESD & CI & Bias & SD & ESD & CI & Bias & SD & ESD & CI  \\
  \hline
& $\beta_1$ & -0.01 & 0.16 & 0.13 & 0.88 & 0.00 & 0.16 & 0.16 & 0.95 & -0.01 & 0.23 & 0.22 & 0.96 \\
1 & $\beta_2$ & -0.06 & 0.19 & 0.21 & 0.98 & -0.01 & 0.19 & 0.25 & 0.99 & -0.05 & 0.30 & 0.36 & 1.00 \\
& $\beta_3$ & 0.15 & 0.38 & 0.29 & 0.82 & 0.08 & 0.41 & 0.38 & 0.92 & 0.18 & 0.71 & 0.73 & 0.97 \\ 
 \hline
& $\beta_1$ & -0.02 & 0.13 & 0.18 & 1.00 & -0.03 & 0.14 & 0.19 & 1.00 & 0.07 & 0.21 & 0.92 & 1.00 \\
2& $\beta_2$ & 0.03 & 0.18 & 0.20 & 0.98 & 0.03 & 0.18 & 0.20 & 0.98 & 0.08 & 0.27 & 1.24 & 0.97 \\
& $\beta_3$ & 0.01 & 0.34 & 0.41 & 1.00 & 0.04 & 0.38 & 0.47 & 1.00 & -0.36 & 0.51 & 5.68 & 0.99 \\ 
 \hline
& $\beta_1$ & 0.04 & 0.12 & 0.12 & 0.98 & 0.00 & 0.12 & 0.14 & 1.00 & 0.10 & 0.13 & 0.17 & 0.97 \\
3& $\beta_2$ & 0.00 & 0.19 & 0.15 & 0.86 & 0.07 & 0.20 & 0.20 & 0.93 & 0.03 & 0.33 & 0.40 & 1.00 \\
& $\beta_3$ & -0.13 & 0.38 & 0.31 & 0.89 & -0.15 & 0.38 & 0.40 & 0.96 & -0.29 & 0.65 & 0.91 & 0.97 \\ 
 \hline
& $\beta_1$ & 0.00 & 0.13 & 0.16 & 1.00 & -0.02 & 0.14 & 0.17 & 1.00 & 0.07 & 0.19 & 1.42 & 1.00 \\
4& $\beta_2$ & 0.00 & 0.21 & 0.25 & 1.00 & 0.02 & 0.21 & 0.25 & 1.00 & 0.11 & 0.30 & 1.94 & 1.00 \\
& $\beta_3$ & 0.02 & 0.29 & 0.35 & 0.98 & 0.04 & 0.29 & 0.43 & 1.00 & -0.41 & 0.60 & 6.58 & 1.00 \\ 
   \hline
\end{tabular}}
\end{table}




\end{document}